\documentclass[conference,square,numbers]{IEEEtran}
\usepackage{hyphenat} %
\PassOptionsToPackage{hyphens}{url}\usepackage{hyperref}
\usepackage{amsmath,amssymb,amsfonts}
\def\BibTeX{{\rm B\kern-.05em{\sc i\kern-.025em b}\kern-.08em
    T\kern-.1667em\lower.7ex\hbox{E}\kern-.125emX}}

\usepackage[utf8]{inputenc}
\usepackage[T1]{fontenc}

\usepackage[]{xcolor}
\usepackage{graphicx}
\usepackage{url}
\usepackage{xspace}
\usepackage{ifthen}
\usepackage{natbib}
\usepackage{enumitem}
\usepackage{textcomp}
\usepackage{tikz}
\usepackage{rotating,array}
\usepackage{footnote}
\usepackage[normalem]{ulem}
\usepackage{pifont} %
\usepackage[binary-units,per-mode=symbol]{siunitx}
\makesavenoteenv{tabular} 
\makesavenoteenv{table} 
\usepackage{cleveref}
\newlist{inlinelist}{enumerate*}{1}
\setlist*[inlinelist,1]{%
  label=(\roman*),
}

\newcommand{\mystyle}[1]{\sloppy{\texttt{{\scalebox{.88}[1.1]{#1}}}}}

\hypersetup{
  colorlinks,
  linkcolor={green!50!black},
  citecolor={red!70!black},
  urlcolor={blue!70!black}
}

\newboolean{showcomments}
\setboolean{showcomments}{false}
\ifthenelse{\boolean{showcomments}}
{ \newcommand{\mynote}[3]{
   \fbox{\bfseries\sffamily\scriptsize#1}
   {\small$\blacktriangleright$\textsf{\emph{\color{#3}{#2}}}$\blacktriangleleft$}}}
{ \newcommand{\mynote}[3]{}}

\newcommand{\vs}[1]{\mynote{Valerio}{#1}{blue}}

\newcommand{\SYS}{\textsc{Tz4Fabric}\xspace}
\newcommand{\sys}{\SYS} %

\date{}

\title{TZ4Fabric: Executing Smart Contracts\\with ARM TrustZone\vspace{-5pt}}
\IEEEspecialpapernotice{\small{(Practical Experience Report)}\vspace{-5pt}}

\author{
Christina Müller\IEEEauthorrefmark{2},
Marcus Brandenburger\IEEEauthorrefmark{3}, 
Christian Cachin\IEEEauthorrefmark{2},\\ 
Pascal Felber\IEEEauthorrefmark{1},
Christian Göttel\IEEEauthorrefmark{1},
Valerio Schiavoni\IEEEauthorrefmark{1}
\\
\IEEEauthorblockA{
\IEEEauthorrefmark{1}Institut d'Informatique, Universit\'e de Neuch\^atel, Switzerland, \texttt{first.last@unine.ch}\\
\IEEEauthorrefmark{3}IBM Research Zurich, Switzerland, \texttt{BUR@zurich.ibm.com}\\
\IEEEauthorrefmark{2}Cryptology and Data Security Research Group, University of Bern, Switzerland, \texttt{first.last@unibe.ch}\\
}
}

\newcommand\copyrighttext{\footnotesize \textcopyright 2020 IEEE.
Personal use of this material is permitted.
Permission from IEEE must be obtained for all other uses, in any current or future media, including reprinting/republishing this material for advertising or promotional purposes, creating new collective works, for resale or redistribution to servers or lists, or reuse of any copyrighted component of this work in other works.
Presented in the \href{https://srds-conference.org/index-real.html}{39th IEEE International Symposium on Reliable Distributed Systems (SRDS '20)}.%
The final version of this paper is available under DOI:
\href{https://doi.org/10.1109/SRDS51746.2020.00011}{10.1109/SRDS51746.2020.00011}}

\newcommand\copyrightnotice{\begin{tikzpicture}[remember picture,overlay]
\node[anchor=south,yshift=10pt,fill=yellow!20] at (current page.south) {\fbox{\parbox{\dimexpr\textwidth-\fboxsep-\fboxrule\relax}{\copyrighttext}}};
\end{tikzpicture}}

\begin{document}

\maketitle
\copyrightnotice
\begin{abstract}
Blockchain technology promises to revolutionize manufacturing industries.  
For example, several supply chain use cases may benefit from transparent asset tracking and automated processes using smart contracts.  
Several real-world deployments exist where the transparency aspect of a blockchain is both an advantage and a disadvantage at the same time.
The exposure of assets and business interaction represent critical risks.
However, there are typically no confidentiality guarantees to protect the smart contract logic as well as the processed data.
Trusted execution environments (TEE) are an emerging technology available in both edge or mobile-grade processors (\emph{e.g.}, ARM TrustZone) and server-grade processors (\emph{e.g.}, Intel SGX).
TEEs shield both code and data from malicious attackers.
This practical experience report presents \sys, an extension of Hyperledger Fabric to leverage ARM TrustZone for the secure execution of smart contracts.
Our design minimizes the trusted computing base executed by avoiding the execution of a whole Hyperledger Fabric node inside the TEE, which continues to run in untrusted environment.
Instead, we restrict it to the execution of only the smart contract.
The \sys prototype exploits the open-source OP-TEE framework, as it supports deployments on cheap low-end devices (\emph{e.g.}, Raspberry Pis). %
Our experimental results highlight the performance trade-off due to the additional security guarantees provided by ARM TrustZone.
\sys will be released as open source.
\end{abstract}
\section{Introduction}
\label{sec:intro}

Industry 4.0~\cite{lasi2014industry} is among the primary pushing factors for the adoption of blockchain technologies.
Beyond the well-known case of \emph{cryptocurrencies}~\cite{miraz2018applications}, several applications are being developed and deployed across diverse domains, including asset trading~\cite{chiu2019blockchain}, insurance claim processing~\cite{raikwar2018blockchain}, cross-border payments~\cite{bitspark,abra}, innovative money lending~\cite{tapscott2017blockchain}, transparent asset tracking along a given supply chain (\emph{e.g.}, farm tracking~\cite{farmerconnect} or trusted food chain~\cite{ibmfoodtrust,tradelens}) and so on.
Typically, all of these scenarios involve a very large and heterogeneous set of mutually untrusted entities which still must collaborate to validate and execute a common (application-specific) transaction.

When deployed over a blockchain infrastructure, the interaction between these entities can be controlled by smart contracts~\cite{2017:BC}, \emph{i.e.}, programs that execute on top of a distributed ledger in a fully decentralized manner~\cite{smartcontracts}. 
The correct execution of the smart contract is enforced by consensus protocols.
Smart contracts can be deployed and executed over ``Internet of things'' (IoT) devices in a verifiable and efficient manner~\cite{2016:SC-IoT,atlam2019intersections}.
Nevertheless, the confidentiality of the contract itself is at stakes, especially in environments where the hardware and software integrity of the device can be compromised by powerful attackers, including compromised operating systems or malicious human operators.

\begin{figure}[!t]  
\centering
\includegraphics[scale=0.65,trim={0 35pt 0 35pt}]{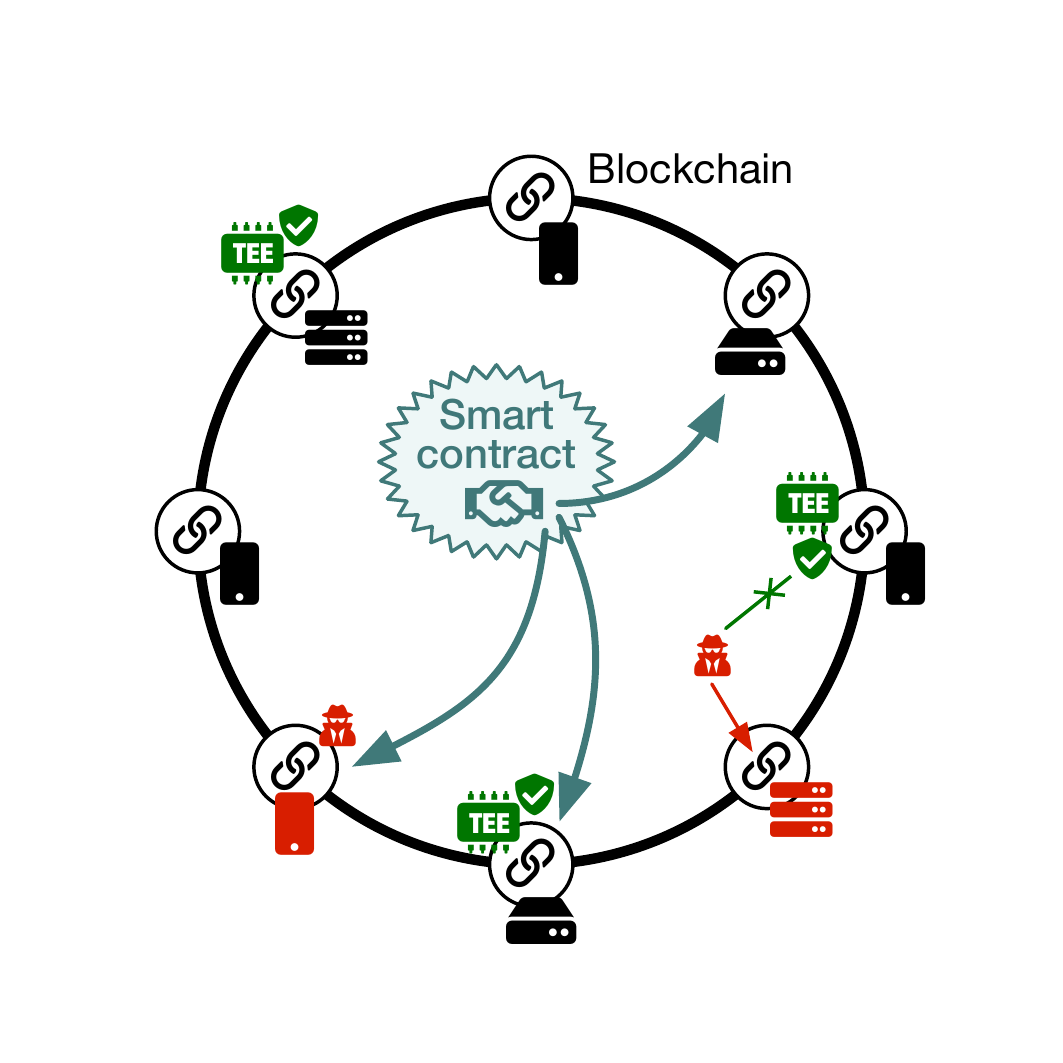}
\caption{Distributed ledger deployment over an heterogeneous set of devices. A subset of them are shielded against malicious attackers using TEEs.}
\label{fig:contract}
\end{figure}

Hardware-based trusted execution environments (TEEs) offer an exciting new opportunity to overcome such limitations.
In a nutshell, TEEs isolate data and code by shielding it from malicious users, compromised system libraries or thwarted operating systems.
Notable examples include Intel software guard extensions (SGX) for server-grade processors~\cite{sgx} and ARM TrustZone~\cite{armtz-doc,armtz-product,ngabonziza2016trustzone} (from here on referred to as ARM TZ) for IoT- and edge-based devices.

Hyperledger Fabric private chaincode (FPC)~\cite{2018:fpc,fpc-online} is an open source project which enables nodes on a channel (a subset of nodes in the blockchain network) the execution of smart contracts inside TEEs, specifically inside Intel SGX enclaves.
Since IoT devices are mostly small, (potentially) battery-powered and embedding low-power processors~\cite{low-power1, low-power2, sgx}, FPC cannot easily be deployed in this context, which greatly hinders its applicability of our target deployment scenarios.

\Cref{fig:contract} depicts our envisioned deployment scenario. 
A heterogeneous and possibly distributed set of nodes joins the blockchain.
A smart contract must be executed across a channel and only few nodes are shielded against attackers.
For example, an attacker could compromise a channel, gain access to its nodes and gather sensitive information.
In order to contain the attack and prevent it from spreading to other channels, smart contracts of some channels could be executed inside a TEE.
Nodes protected by TEEs are shielded against malicious attacks, including a compromised operating system or an attacker with physical access to the device. 
Especially nodes that are deployed as IoT devices in ``the wild'' are prone to such kinds of attacks.
In the case of TEE-enabled nodes, the processor package is the security perimeter.

In this paper, we present the design, implementation and evaluation of \sys, a prototype for Hyperledger Fabric~\cite{fabric-doc} chaincode execution that integrates with ARM TZ.
We discuss in particular the challenges that arise from executing smart contracts in an embedded TEE much less powerful than Intel SGX, as well as the limitations that result from \sys's design and the underlying platform.
As we detail further (\Cref{sec:archi}), \sys architecture is inspired by FPC, but it isolates the TEE component so that only the smart contract execution can be offloaded to it.
This design allows not only to leverage ARM TZ, but also to possibly extend \sys to future (yet unreleased) TEEs, by leveraging our modular architecture.
We believe this modularity represents a major differentiation, but can also be seen as a supplement for FPC's design.

This paper is organized as follows.
In \Cref{sec:background}, we present background on blockchain and Hyperledger Fabric, as well as ARM TZ and the OP-TEE runtime.
\Cref{sec:threat} describes \sys's threat model.
We then describe the architecture of our system in \Cref{sec:archi} and elaborate on its most important implementation details in \Cref{sec:impl}.
\Cref{sec:eval} presents the evaluation of our prototype, reporting on performance as well as energy-related results.
We finally discuss related work in \Cref{sec:rw} and conclude with some open challenges in \Cref{sec:conc}.
\section{Background}
\label{sec:background}

\begin{figure}[!t]  
\centering
\includegraphics[scale=0.68]{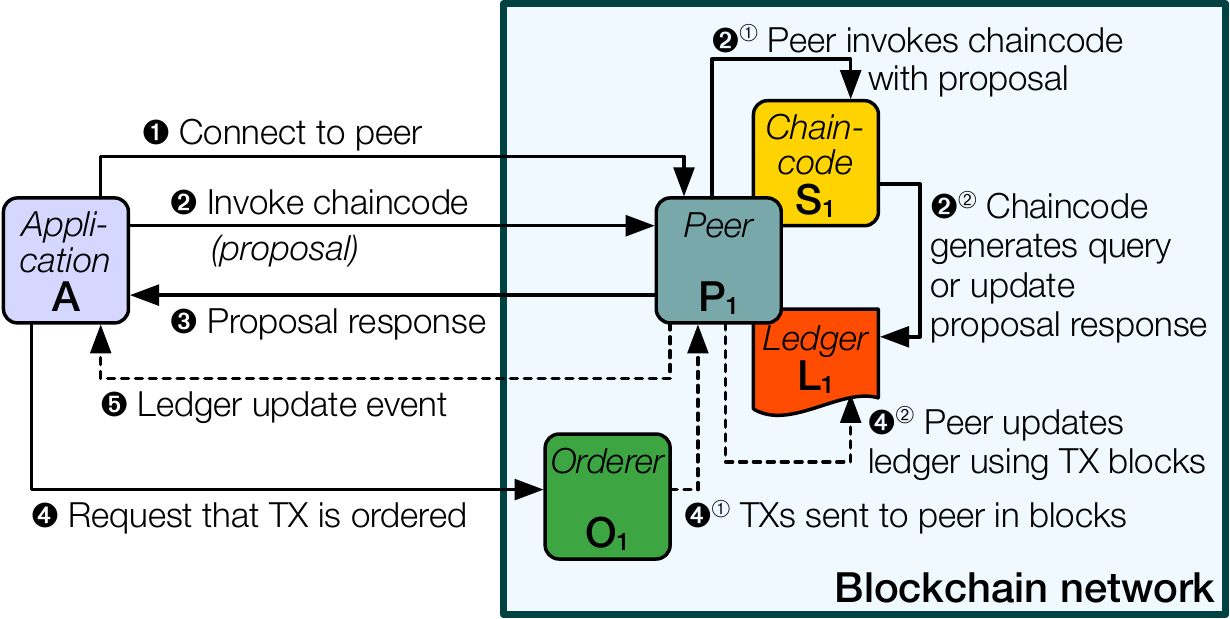}
\caption{Architecture of Hyperledger Fabric.}
\label{fig:hfarc}
\end{figure}

\textbf{Blockchain and Smart-contracts.}
A blockchain is a type of distributed ledger~\cite{2017:BC}.
It records data (\emph{i.e.}, transactions) in a decentralized way.
The data is appended in blocks and connected (``chained'') via hashes.
Each transaction is signed by the party it was invoked.
Before appending transactions, nodes of the blockchain network must validate and agree on a unique order of these transactions.
The latter can be achieved via some consensus mechanism~\cite{badertscher2017bitcoin,garay2015bitcoin,bentov2016snow,david2018ouroboros}.
Thanks to its characteristics, a blockchain guarantees availability, transparency, immutability and integrity of the stored data.
Data privacy and scalability are rather limited in blockchains.%
We distinguish between the following two types of blockchains: permissionless (also called public) and permissioned~\cite{2017:CV}.
In a public blockchain (\emph{e.g.}, Bitcoin~\cite{bitcoin-online} and Ethereum~\cite{ethereum-online}), anyone can participate in the network (read data from the blockchain, invoke transactions, validate transactions, \emph{etc.}), whereas in a permissioned blockchain (\emph{e.g.}, Ripple~\cite{ripple-online} and Hyperledger Fabric~\cite{2018:fabric-paper}), the access of the network is restricted and entities are known.
The concept of smart contracts was described first by Nick Szabo~\cite{smartcontracts}: \emph{A smart contract is a computerized transaction protocol that executes the terms of a contract.}
With the emergence of blockchain technology, this idea was put into practice.
A smart contract is an agreement translated into program code and stored in a blockchain network~\cite{18:sc, dzone:sc}.
It is automatically executed when the defined conditions are met.
As they are integrated into a blockchain, smart contracts inherit from its features of availability, transparency, immutability and integrity~\cite{2017:BC}.
Furthermore, they are efficient, reduce cost and save time bypassing any third party.
However, the execution of smart contracts present several challenges, \emph{e.g.}, it is not trivial to guarantee data privacy, avoid bugs in the contract code or protect from attacks, as it happened to Ethereum~\cite{17:eth-attacks}.
\textbf{Hyperledger Fabric} (HF)~\cite{2018:fabric-paper,fabric-doc} is a permissioned blockchain supporting smart contracts.
It belongs to the Hyperledger~\cite{hyperledger} open source project that includes different frameworks and tools related to blockchain technologies.
 In HF, a smart contract is called \emph{chaincode}.
Currently, developers can choose among three general-purpose programming languages (Go, Java and Node.js) to implement chaincodes.
To facilitate deployments, HF uses Docker containers.
There exist three types of nodes in a HF network: clients, peers and orderers.
\Cref{fig:hfarc} presents the high-level architecture and workflow of HF.
Before a chaincode function can get called (invoked), it must be installed (stored on the file system) and instantiated at the peer.
A client (application) sends a request (transaction proposal) to a peer in order to invoke a chaincode function (\Cref{fig:hfarc}-\ding{202} and \Cref{fig:hfarc}-\ding{203}). %
In a first phase, \emph{i.e.}, execution or \emph{endorsement}, the peer executes the called chaincode function (\Cref{fig:hfarc}-\ding{203}\textsuperscript{\ding{192}} and \Cref{fig:hfarc}-\ding{203}\textsuperscript{\ding{193}})
and sends a transaction response %
back to the client (\Cref{fig:hfarc}-\ding{204}). %
The transaction response is signed by the peer and contains the execution response message, as well as the readset and writeset.
The readset represents all values of the keys a peer has queried from the ledger via \mystyle{GetState} during the execution.
The writeset contains all key-value pair updates a peer has generated via \mystyle{PutState}.
When the client has collected enough transaction responses as defined by the so called endorsement policy it sends them to an orderer (\Cref{fig:hfarc}-\ding{205}). %
The orderer puts the transaction responses into blocks and disseminates them to all peers (\Cref{fig:hfarc}-\ding{205}\textsuperscript{\ding{192}}). %
This is the \emph{ordering} phase.
In the third phase, the \emph{validation} phase, each peer checks if the endorsement policy is satisfied and if there is no read-write conflict between the different transactions.
Finally, they put the transaction on the ledger (\Cref{fig:hfarc}-\ding{205}\textsuperscript{\ding{193}}). %
The ledger has two components: a blockchain and a world state.
The world state is a pluggable database, to store and efficiently retrieve later on the current values of the keys in the blockchain.

\textbf{ARM TrustZone} provides hardware components for enabling TEEs on ARM processors~\cite{armtz-doc, trust-exec, armtz-product}.
OP-TEE~\cite{optee-doc} is a popular open source runtime with native support for ARM TZ.
OP-TEE follows the TEE architecture and API standardized by GlobalPlatform (GP)~\cite{gp-online}.
ARM TZ enables a single TEE---called secure world---per system~\cite{tz-explained}.
The other part of the system is called normal world.
The processor can be in one of two security states: secure (for the secure world) and non-secure (for the normal world).
Switching happens via a secure monitor call (SMC).
System resources are strictly isolated: the normal world cannot access the resources (\emph{e.g.}, memory, peripherals, \emph{etc.}) reserved for the secure world.
During the bootstrap of the secure world, a chain of trust is established and there is an integrity check of the secure world software images---a process called secure boot.
Additional details regarding OP-TEE framework are given while discussing \sys implementation in \Cref{sec:impl}.
\section{Threat Model}
\label{sec:threat}
For our thread model we consider a powerful attacker with administrative rights as well as physical access to all nodes supporting ARM TZ.
There, we rely on the protection mechanisms offered by the TEE to shield the privacy of the smart-contract code, as well as its privacy upon execution.
We further assume that the operating system and the user space in the normal world cannot be trusted.
However, we do assume that the TEE, which includes bootloader, firmware, OP-TEE, and the secure monitor are trusted.
The HF blockchain network can run different consensus algorithms, among which PBFT~\cite{pbft}, and can therefore tolerate Byzantine failures.

Since ARM TZ does not natively support remote attestation mechanisms, an attacker might try to compromise the contract before it is executed.
The chaincode is stored in the normal world as a trusted application (TA) and signed with the build key.
In addition, it is possible to encrypt the TA to add further security measures.
Without the build key an attacker cannot easily tamper with the chaincode, as its signature is verified in the secure world before execution.
Furthermore, this attack can be mitigated by integrating schemes such as Fides~\cite{Prunster2019}.
Notice that ARM TZ does not provide integrity protection at runtime, hence \sys cannot easily prevent unnoticed malicious corruptions of the chaincode.

While recent side-channel attacks~\cite{10.1145/3319535.3354197,10.1145/3319535.3354201} have been unveiled against ARM TZ, and mitigations continue to be released~\cite{arm-speculation-barrier}, we consider them out of the scope of this work, as also mentioned in~\cite{armtz-doc}.
Mitigations will eventually be released by means of architectural microcode updates (as happened for Intel SGX to mitigate Spectre/Meltdown attacks~\cite{lipp2018meltdown}).
\section{\sys Architecture}
\label{sec:archi}
ARM TZ-enabled devices are typically low-power embedded devices of which many are battery powered.
Hence, these devices have to get by with limited resources in terms of persistent and volatile memory.
Given these limitations, it is oftentimes challenging to port applications or systems that are being used in a desktop or sever environment -- in particular if these tend to have many dependencies.
In our case, \sys depends on packages such as Go programming language (golang) environment, HF, OP-TEE, as well as gRPC~\cite{grpc}.
Not only do these packages (and their dependencies) require several hundreds of megabyte of persistent memory, but also do they need a few hundred megabytes of volatile memory at runtime.
While there are IoT devices that can satisfy these requirements (\emph{e.g.}, Raspberry Pi models), many cannot.
For this reason we have settled for an approach where we decouple large system components (\emph{i.e.}, HF and Docker) from lightweight security-relevant components (\emph{i.e.}, chaincode) by means of a proxy.

The design of \sys is inspired by the architecture of FPC~\cite{2018:fpc,fpc-online}.
In our design, sensitive information can be contained in the chaincode operation and its response, which have to be shielded from malicious attacks.
We therefore offload the chaincode to a TEE-enabled embedded device.
To facilitate our design, we divide \sys into three major components: \emph{(1)} a wrapper that resides as a chaincode at a peer, \emph{(2)} a proxy and \emph{(3)} the chaincode itself.
The wrapper communicates with the proxy via gRPC.
The proxy and the chaincode run on an ARM-based environment with TrustZone.
In particular, the proxy resides in the normal world, whereas the chaincode is located in the secure world.
Our design leverages the OP-TEE framework to drive the interaction between the proxy and the chaincode.
The wrapper is installed and instantiated as chaincode and is used as an interface towards the peer and the ledger.
It forwards incoming invocations from the clients to the chaincode in the secure world, handles the communication towards the ledger and sends transaction responses back to the peer.
The proxy acts as an intermediary and forwards the calls between the wrapper and the chaincode.
It is responsible for context switching from the normal world to the secure world.
A chaincode implements the blockchain application logic and is invoked by the clients via the wrapper.
During execution, a chaincode has access to the ledger via \mystyle{getState} and \mystyle{putState} commands through the proxy.

Our design allows us to instantiate and run multiple chaincodes in a single secure world, which provides isolation from the normal word, \emph{i.e.}, the operating system does not have direct access to the chaincode resources and can only interact with a chaincode through the \mystyle{proxy}.
However, as all chaincodes reside in the same secure world, there are no isolation guarantees among the chaincodes, something that is instead provided by FPC as a result of the ability to run multiple SGX enclaves on the same processor.

Our prototype focuses on the inclusion of ARM TZ in HF.
This means that we have not implemented a mechanism in order to replicate chaincodes in our network.
The principal reason for this choice lies in the execution of chaincodes inside ARM TZ.
Chaincodes are executed in the prototype as TAs (see \Cref{sec:impl}) inside OP-TEE.
This implies that the chaincode is compiled using the same build system as the OP-TEE running on the machine and that the TA is signed with the original build key and optionally encrypted.
Therefore, every chaincode would have to be compiled to a TA and installed on the target machine, before it could be invoked.
Alternatively, implementing a virtual machine as TA inside ARM TZ that is capable to execute smart contracts (\emph{e.g.}, Solidity~\cite{solidity}) would overcome the previously explained restriction.
Chaincodes could then be replicated on the target machine and directly be invoked without having to go through the process of generating a TA.
In the prototype we have generated a TA for the chaincode and we have deployed it on all ARM TZ-enabled \mystyle{proxies}.
As mentioned before, there is also the possibility to use smart contracts which are written in domain-specific languages (DSL) (\emph{e.g.}, Solidity) and that are executed by a virtual machine (\emph{e.g.}, Ethereum Virtual Machine).
Since our goal is to shield the chaincode with a TEE, this would also require that the virtual machines executing the chaincode is shielded as well.
Such an approach would inevitably increase the trusted computing base (TCB) and take up some of the already limited resources on embedded devices.
On top of that our design would strongly depend on the design of the DSL and the potential flaws that come with it.

Another aspect to consider is that the ledger might hold sensitive information, which could be shielded by the approach used in FPC.
However, for our prototype we did not shield the ledger using Intel SGX.
\section{Implementation details}
\label{sec:impl}

\begin{figure}[!t]  
\centering
\includegraphics[scale=0.65]{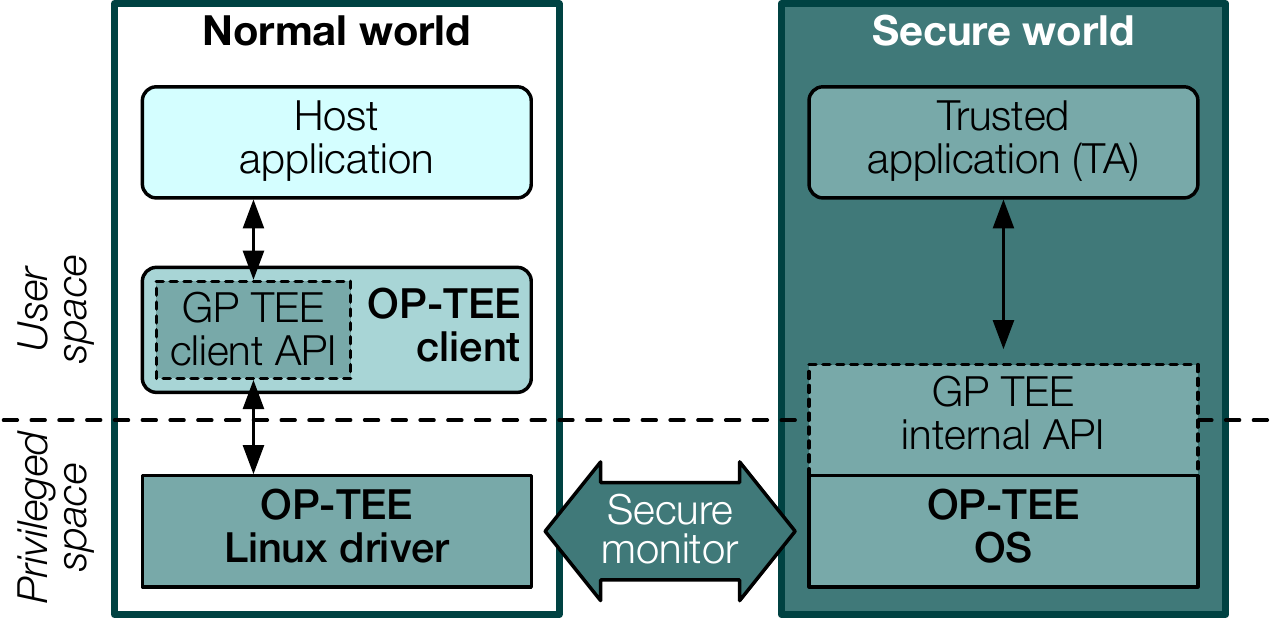}
\caption{Architecture of OP-TEE (trusted parts in darker shades of green).}
\label{fig:optee-architecture}
\end{figure}

Our implementation leverages the OP-TEE framework, depicted in \Cref{fig:optee-architecture}.
OP-TEE contains the following components: OP-TEE Client, OP-TEE Linux driver and OP-TEE OS.

The OS of the normal world is also referred to as \emph{rich execution environment} (REE).
The OP-TEE Linux driver provides the driver for the normal world.
An application running inside the normal world is referred to as host application.
The TEE client API and the TEE internal API enable the communication between a host application and an application of the secure world, \emph{i.e.}, the TA.
Both APIs are defined by GP~\cite{client-api, internal-api}: the TEE client API is implemented by the OP-TEE client component, the TEE internal API is implemented by the OP-TEE OS. 

Interactions between the different components of the system happens as follows (see also \Cref{fig:latency-breakdown-setup}).
Before initiating any communication, the host application establishes a connection towards the secure world (\mystyle{TEEC\_InitializeContext}) and opens a session towards the TA (\mystyle{TEEC\_OpenSession}) with the unique identifier of the TA (\texttt{UUID}) as parameter.
Then, the host application can call functions of the TA with the TEE client APIs \mystyle{TEEC\_InvokeCommand}, as this allows to pass data between the host application and the TA via shared memory reference or by value.

Once the host application has finished communication with the TA, it needs to close the session (\mystyle{TEEC\_CloseSession}) and finalize the context (\mystyle{TEEC\_FinalizeContext}) to release any allocated resources. 

In our implementation, all \sys components are deployed within containers.
Since there is no native Docker support in the normal world (untrusted part) of OP-TEE, the peer is decoupled from the ARM TZ node.
Communication between a \textit{chaincode\_wrapper} at the peer (\sys specific wrapper around the actual chaincode) and a \textit{chaincode} inside the secure world is enabled via gRPC remote procedure calls and through the API provided by OP-TEE between normal and secure world.

Finally, it is worth mentioning that the availability of remote attestation is one of the most important difference between Intel SGX and ARM TZ with OP-TEE, where the former natively supports it~\cite{14-18:ra, optee-no-att, sgx}.
While this represents one of the main limitation of \sys compared to its SGX-based FPC counterpart, we highlight in \Cref{sec:threat} possible workarounds.
The code of \sys is released as open source and available from \url{https://github.com/piachristel/open-source-fabric-optee-chaincode}.
\section{Evaluation}
\label{sec:eval}

\begin{figure*}[!t] %
	\centering	
        \includegraphics[width=0.90\linewidth]{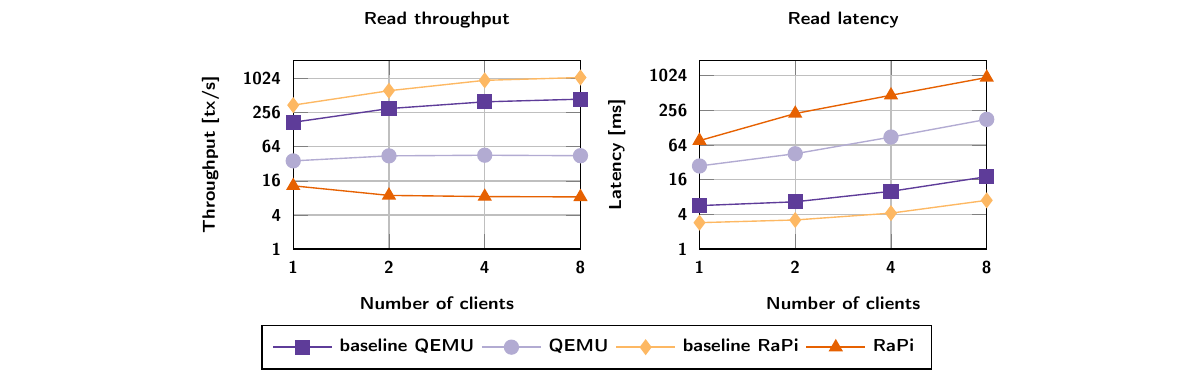}
	\caption{Read throughput and read latency of the coffee tracking chaincode for increasing numbers of clients.}%
	\label{fig:throughput-latency-foc}
\end{figure*}
This section presents the experimental evaluation of the \sys prototype.
Our main goal is to show the performance trade-offs of the system.
In particular, we investigate the throughput and latency impact of our implementation.
Further, to clarify the performance penalties of the system, we conduct a latency breakdown to identify the bottleneck in the system, as well as energy measurements.

\vspace{-4pt}
\subsection{Experimental settings}
\label{sec:settings}
\vspace{-2pt}
Each experiment is executed with OP-TEE~(v3.8) using QEMU for ARM TZ emulation and with OP-TEE running on the Raspberry Pi.
We use a set of 27 server machines with two Intel Xeon CPU L5420 at \SI{2.5}{\giga\hertz} and \SI{8}{\giga\byte} of RAM.
The \mystyle{orderer}, 8 \mystyle{wrapper} and 8 \mystyle{proxy} (QEMU instance) each run on their own server.
The Raspberry Pi model 3B+ embeds ARM TZ (ARMv8-A). 
We employ the Linux CPUFreq powersave governor on all Raspberry Pi.
\vs{@christian: add which CPU governor is used}
Hence, we emulate the same processor using QEMU.
Specifically, QEMU emulates a Cortex-A53 CPU with symmetric multiprocessing (SMP) set to 4 cores, and the number of trusted threads in OP-TEE is set to 4 in order to have an emulated environment to closely mimic the one found in the physical Raspberry Pi 3B+.
Clients invoke transactions from 8 dedicated machines.

We use two auxiliary machines in order to record the power consumption of the different types of nodes that form our HF network.
We rely on a LINDY iPower Control 2x6M power distribution unit (PDU) to monitor energy consumptions.
Due to the limited number of available ports, we restrict the measurements the following nodes: the \mystyle{orderer}, 3 \mystyle{wrapper}, and 3 \mystyle{proxy} (QEMU instances).
The 8 Raspberry Pi 3B+ are all equipped with a Raspberry Pi
PoE-HAT\footnote{\url{https://www.raspberrypi.org/products/poe-hat/}, last access: May 29 2020} and make use of 802.3.af Power-over-Ethernet~\cite{mendelson2004all}.
They are connected to a Ubiquiti Networks UniFi USW-48P-750 switch, which also allows to directly query their power consumption.
The nodes are connected via 1 Gbit/s switched network and run Ubuntu~18.04~LTS, except for the client machines which run Ubuntu~20.04~LTS.
The clocks of the auxiliary and the client machines are synchronized using NTP, which is necessary in order to relate the power measurements to the benchmark measurements in our setup.
We implemented and deploy in our benchmarks a simple chaincode toy-example, \emph{e.g.}, tracking the coffee consumption in an office.
Clients submit transactions to track their coffee consumption and query the current coffee statistics.
The chaincode and the \mystyle{proxy} are written in ISO C and C++, whereas the clients and the \mystyle{wrapper} are written in golang.
Our coffee tracking chaincode is simple enough to conduct a read/write benchmark on the HF blockchain network.
\vs{should we say why this example is representative? baseline numbers?}
Each client machine runs a peer. 
We exploit its capability to connect to a remote peer in the \sys network.
Clients repeatedly invoke transactions on the their local peer which get forwarded to the HF network peer.
The communication between the \mystyle{wrapper} and the \mystyle{proxy} uses gRPC~(v.1.28.1).

The execution time of a client measurement begins before the client invokes the first transaction and terminates after the client has received the corresponding response from the last transaction.
Furthermore, for each client we record the number of invoked transactions.
We support experimental reproducibility, and detailed instructions to setup and reproduce these experiments are given at \sys git repository (\Cref{sec:impl}).

\subsection{TCB Size}
\label{sec:tcb}
The TCB consists of all components residing in the secure world.
With our design we limit this to the chaincode and other components (\emph{i.e.}, \mystyle{wrapper}, \mystyle{proxy}).
The peers are considered to be untrusted and thus run outside the secure world.
In total, our prototype contains $\sim$\num{22000} lines of untrusted C++ code (most of it generated by gRPC) and only $\sim$\num{400} lines of trusted C code.
Additionally, the OP-TEE OS itself consists of $\sim$\num{233600} lines of C and assembler code, as well as the Trusted Firmware-A which sums to $\sim$\num{31400} lines of C and assembler code.
For OP-TEE OS and Trusted Firmware-A we considered only the platform-specific source code.

In contrast, the peer consists of $\sim$\num{80000} lines of Go code~(v1.4.1).
This shows that by only executing the chaincode itself inside the TEE and not the entire peer, we can drastically minimize the TCB and thereby reduce the attack surface.

\subsection{Throughput and Latency Impact}
\label{sec:tputimpact}

We begin with measuring the throughput and latency of \sys, by deploying and running the coffee tracking chaincode.
We perform multiple rounds with up to 8 clients.
In every round the clients repeatedly invoke transactions over at least 30 seconds.
We expect the throughput to increase with increasing number of clients, while latency to be stable until the number of clients matches the CPU cores in our machine.

A preliminary evaluation using a single peer and a single Raspberry Pi 3B has otherwise shown that the throughput only slightly increases, by approximately \SI{1.2}{\times}, from~1 to~2 clients and then already stagnates.

To investigate this unexpected result and identify the root cause of the bottlenecks, we conducted the experiment with different scenarios.
Our baseline consists of all components (\emph{i.e.}, chaincode, clients, \emph{etc.}) running in the normal (untrusted) world on the same machine.
This scenario does not benefit from the security properties introduced by ARM TZ, and it does not use the gRPC messaging library.
In a second scenario we place the chaincode in the secure world, but run the components on the same machine.
Finally, we locate the clients and the \mystyle{wrapper} on a dedicated machine, and \mystyle{proxy} and chaincode runs with QEMU and on a Raspberry~Pi.

\Cref{fig:throughput-latency-foc} summaries the throughput and latency for the baseline scenario and the final scenario.
   
In the baseline scenario, we look into the impact of gRPC.
We tried different configurations and the best results are achieved with \mystyle{NUM\_CQS} set to half the number of cores, and \mystyle{MIN\_POLLERS} and \mystyle{MAX\_POLLERS} with the default values.
Using these settings, throughput increases by \SI{1.9}{\times} for 2 clients, \SI{2.5}{\times} for 4 clients and \SI{2.8}{\times} for 8 clients.
As these factors are far above those observed in our preliminary experiment, we can therefore rule out gRPC as the bottleneck.
   
In the second scenario, we notice that the throughput only increases by \SI{1.3}{\times} for 2 clients, \SI{1.4}{\times} for 4 clients and \SI{1.5}{\times} for 8 clients.
These factors are slightly above our observation during our preliminary experiment, but still well below the baseline case.
We therefore conclude that the bottleneck mainly originates from the secure world calls. 

\begin{figure}[!t]
	\centering
	\includegraphics[scale=0.52]{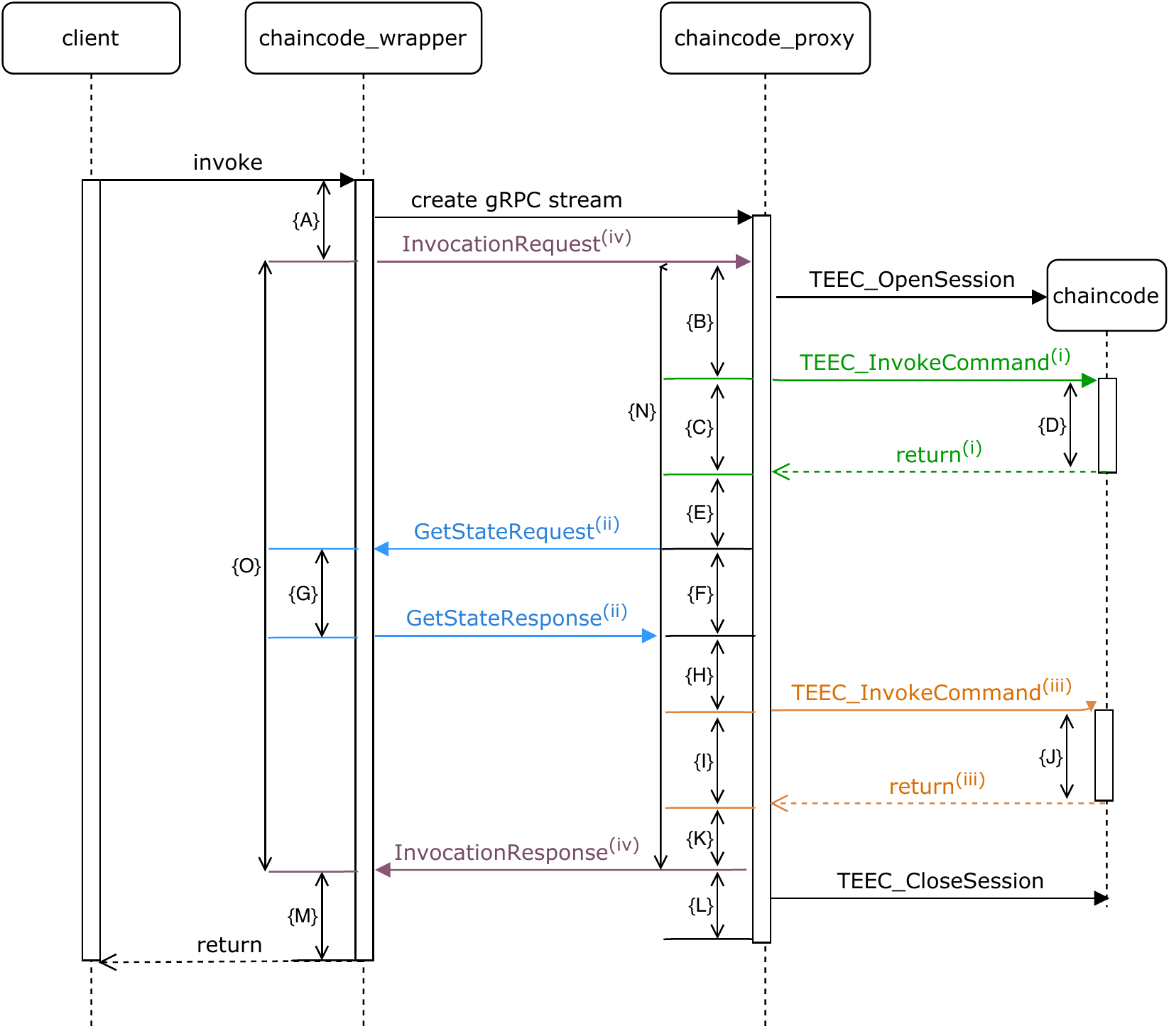}
	\caption{Latency breakdown for \sys.}
	\label{fig:latency-breakdown-setup}
\end{figure}

In the final scenario we observe that the throughput does not improve with an increasing number of clients in case of the Raspberry Pi.
In contrast with QEMU, there is an increase of  \SI{1.2}{\times} from 1 to 2 clients.
For more clients, throughput stagnates.
This is due to the bottleneck we have already observed in the second scenario.
In contrast, the throughput of the baseline increases up to 8 clients.
With the \mystyle{proxy} and the \mystyle{wrapper} running on the Raspberry Pi, we can increase throughput by \SI{3.1}{\times} for 8 clients compared to 1 client.
For QEMU, it is increased by \SI{2.6}{\times} from 1 to 8 clients.

Considering the experiment on the Raspberry Pi for 1 client, the throughput of the baseline is about \SI{27}{\times} higher compared to original \sys.
In case of 8 clients, we even have an increase of about \SI{130}{\times}.
For QEMU, the difference is less extreme: the baseline throughput is between \SI{5}{\times} to \SI{10}{\times} higher compared to the original \sys.
The observations show that the execution of the \mystyle{chaincode} inside the secure world comes with non-negligible cost in terms of throughput.

\subsection{Latency Breakdown}

Next, we provide a detailed breakdown of the latency overheads incurred by \sys.
We divide the execution latency in several phases as shown in \Cref{fig:latency-breakdown-setup}:

\emph{A} initializes the \mystyle{wrapper};
\emph{B} initializes the \mystyle{proxy} and includes \mystyle{TEEC\_OpenSession} to allocate memory;
\emph{D} prepares \mystyle{GetState} commands in the \mystyle{chaincode};
\emph{E} forwards the command via \mystyle{GetStateRequest} to the \mystyle{wrapper};
\emph{G} processes \mystyle{GetStateRequest};
\emph{H} post-processes \mystyle{GetStateResponse} in the \mystyle{proxy};
\emph{J} executes the response message in the \mystyle{chaincode};
\emph{K} concludes the \mystyle{chaincode} execution via a \mystyle{InvocationResponse};
\emph{L} releases the resources, for closing the session (\mystyle{TEEC\_CloseSession}) and for finalizing the context (\mystyle{TEEC\_FinalizeContext}) at the \mystyle{proxy};
\emph{M} finalizes execution at the \mystyle{wrapper}.
The phases \emph{C}, \emph{I}, \emph{N}, and \emph{O} are measured for calculating the round-trip times.

We perform this experiment with the same settings as in the original \sys experiment. 

For both settings with QEMU and the Raspberry Pi, phase \emph{B} contributes up $65$-$75\%$ of the total time.
We observe that the vast majority of phase B is needed for \mystyle{TEEC\_OpenSession}.
We leverage this information to identify throughput bottlenecks observed throughout our evaluation.

The measured latency for the phases \emph{D}, \emph{E}, \emph{G}, \emph{H}, \emph{J}, \emph{K} and \emph{M} are smaller than $1$~ms, and overall contribute to less than $1\%$ of the total latency.

\begin{figure*}[ht]
  \centering
  \includegraphics[width=0.90\linewidth]{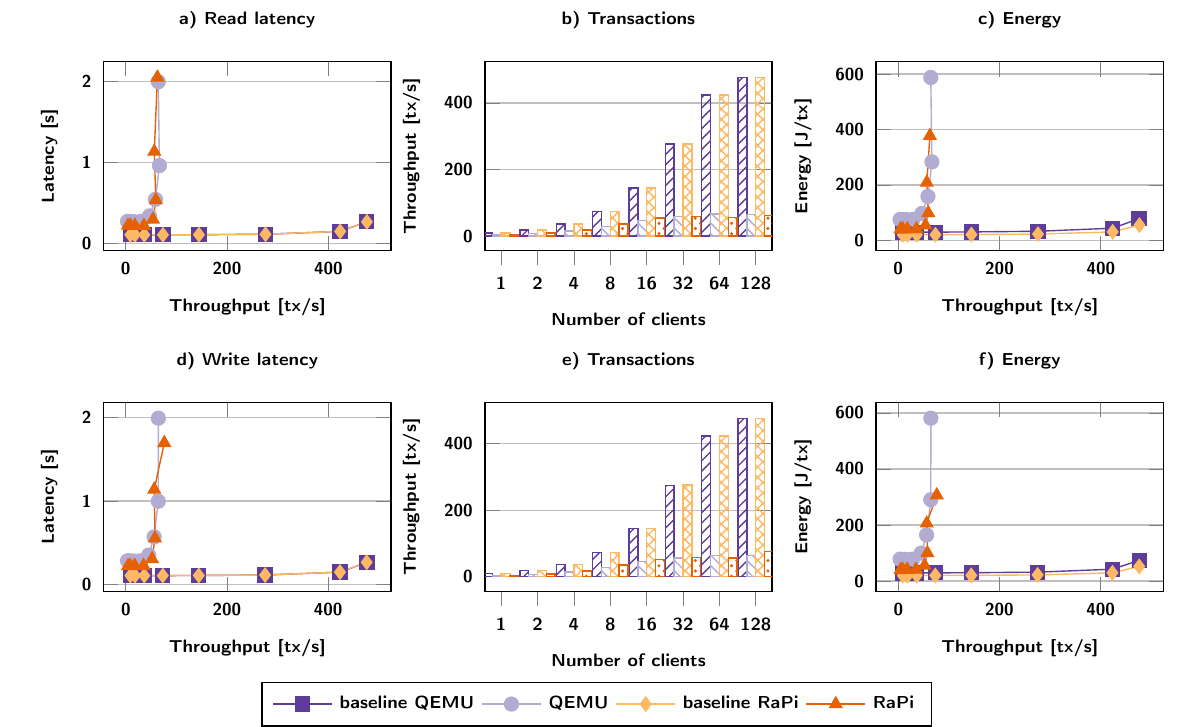}
  \caption{Throughput-latency, transactions and transaction energy for read/write invocations. Top row are read transactions, bottom row are write transactions.\label{fig:bench}}
\end{figure*}

\subsection{Throughput-Latency Benchmark}
In this benchmark we evaluate the coffee tracking chaincode on the setup described in \Cref{sec:settings}.
We perform multiple rounds with up to 128 clients.
In every round the clients repeatedly invoke read or write transactions over a duration of 5 minutes.
Clients are distributed evenly among the 8 available peers.
We use the same naming convention for the scenarios introduced in \Cref{sec:tputimpact} with the exception that all components (excluding clients) run on their own server.

The throughput-latency plot is shown in \Cref{fig:bench}.
We observe that for both read and write transactions, the baselines are very similar to each other.
In the baseline scenario, the chaincode is executed on the \mystyle{wrapper} in the normal world and not shielded by ARM TZ.
There is also no communication going on from the \mystyle{wrapper} to the \mystyle{proxy} since this scenario does not make use of gRPC.
Therefore, the setup for these two baselines is identical, in line with our measurements.

Notice that our benchmark is not saturating the system with the baseline settings.  %
We consistently observe across transactions that the ARM TZ-enabled
environments start saturating already with 8 clients around \SI{65}{tx\per\second}, which corresponds to a single client per \mystyle{wrapper}-\mystyle{proxy} pair.
This observation was made earlier during the preliminary evaluation in \Cref{sec:tputimpact}.
Client transactions on the chaincode are essentially invoked sequentially.
One reason for the low throughput is the substantial overhead due to shielding TAs with ARM TZ.
There is also potential to improve the load times for TAs with OP-TEE by caching their contexts and session to improve reusability.
We intend to explore these optimizations in future work.

Using a shielded QEMU instance (emulating ARM TZ),  the latency increases by a factor of $2.5$ for reads and by a factor of $2.6$ for writes respectively, before reaching the saturation point.
On the Raspberry Pi the latency increases by a factor of $2.0$ for reads and by a factor of $2.1$ for writes before reaching the saturation point.
While the throughput and latency of read transactions is better by a factor of $1.04$ than write transactions with QEMU, the throughput and latency of read transaction improves by a factor of $1.02$ than write transactions on the Raspberry Pi.

\begin{figure*}[ht]
  \centering
  \includegraphics[width=0.90\linewidth]{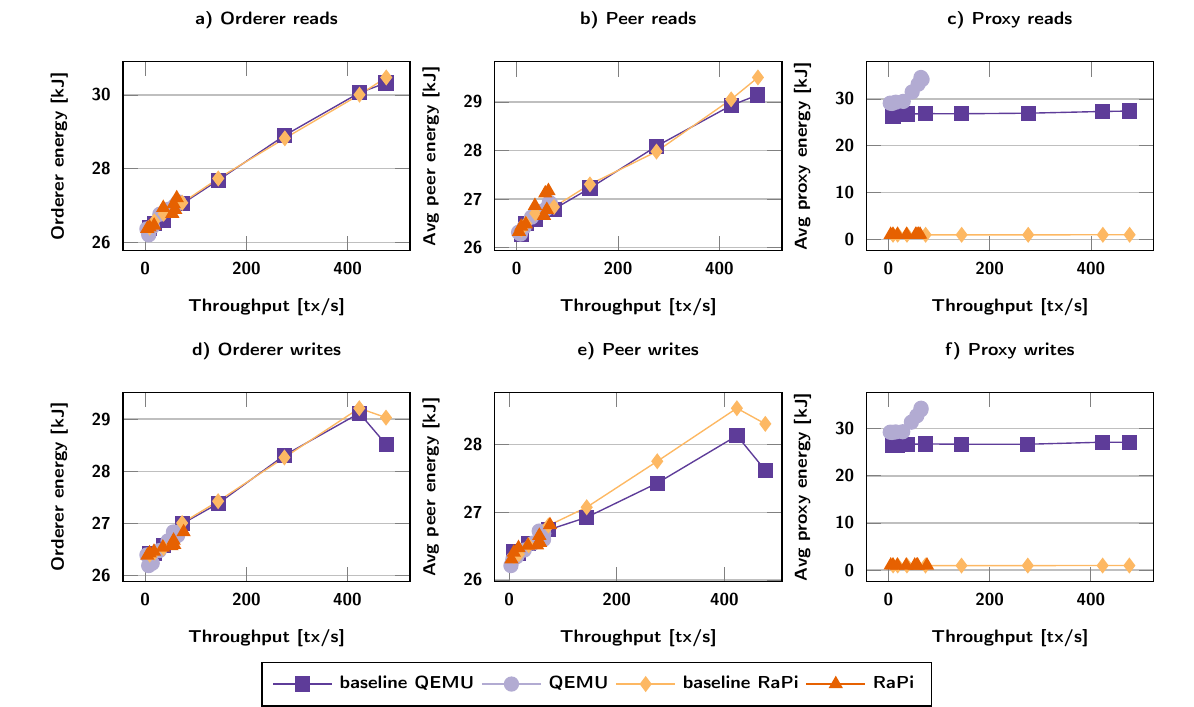}
  \caption{Energy consumption of nodes in the HF network. Top row are read transactions, bottom row are write transactions.\label{fig:energy}}
\end{figure*}

\subsection{Energy consumption}
Finally, we conclude our experimental evaluation with a study of the power footprint of \sys.
We achieve this by recording the power consumption of the nodes running the HF network with the coffee tracking chaincode with \sys.
We expect the execution to be faster and require less energy per transaction when running OP-TEE on the Raspberry Pi as compared to emulating ARM TZ with QEMU.

The energy consumption of the \mystyle{orderer}, \mystyle{wrapper}, and \mystyle{proxy} are depicted in \Cref{fig:energy}, and we distinguish between read (top) and write (bottom) transactions.
We observe that the energy consumption across the nodes during the entire benchmark is rather stable, only slightly increasing when the number of clients grow.
There are two exceptions occurring under different situations for 16 clients: \emph{(1)} the \mystyle{orderer} and \mystyle{wrapper} energy for the baselines rise (\Cref{fig:energy}-[a,b,d,e]) as well as \emph{(2)} the energy of the \mystyle{proxy} (\Cref{fig:energy}-[c,f]).

The first exception is an indication of saturation of the HF network.
This is further supported by the decline of average transactions per client in the network and the fact that the system starts approaching the saturation point as shown in \Cref{fig:bench}.
Both the \mystyle{orderer} and \mystyle{wrapper} nodes begin to struggle for resources under the load of the increasing number of clients.

The second exception is a result of the ARM TZ world-switching overhead when a chaincode is invoked.
As the load increases the performance governor has to increase the frequency and voltage of the CPU over longer intervals.
Although barely visible, the Raspberry Pi is experiencing the same issue.
This is better highlighted in the energy per transaction (see \Cref{fig:bench}), where the energy increases proportionally to the QEMU instance.
This steadily raises the energy consumption of the \mystyle{proxy} machines by a factor of $1.25$ with QEMU and by a factor of $1.05$ on a Raspberry Pi beyond the saturation point.
In terms of energy there is no significant difference between read and write transactions on both platforms.
\section{Related Work}
\label{sec:rw}

\paragraph*{Smart contract execution with a TEE}
Apart from FPC, there are some other works about confidential smart contract execution with a TEE.
Confidential consortium framework (CCF)~\cite{ccf-doc}, Ekiden~\cite{18:ekiden}, ShadowEth~\cite{18:shadoweth} and private data objects (PDOs)~\cite{18:pdo}, just to mention a few recent ones.
In contrast to \sys, they all use Intel SGX as underlying TEE technology for a (prototype) implementation.
The authors of Ekiden state that their technology may use any TEE which is similar to Intel SGX and supports attestation.
Furthermore, SGX can provide means to mitigate rollback attacks~\cite{9049585}, for which the necessary hardware support might not be provided in ARM TZ.

\paragraph*{Confidentiality in context of blockchain and IoT}
In \sys, we are concerned with confidentiality of smart contracts (their logic) and data in context of IoT networks.
Some other research efforts are concerned with the confidentiality of data produced and processed by IoT devices and stored on the blockchain.

\paragraph*{Trust for data generated by IoT devices}
AnyLedger~\cite{anyledger} is a platform for connecting physical devices to the blockchain.
The key feature is an ARM TZ-based wallet for IoT devices.
Key generation, private key storage and signing process of (smart contract) transactions are all placed inside the secure world of ARM TZ.
Hence, AnyLedger wallet guarantees that IoT data hash / address (linking to the interplanetary file system IPFS) placed on the blockchain is integrity protected and authenticated.
Furthermore, data stored on the IPFS is encrypted.
AnyLedger is pluggable to any blockchain technology (for example Ethereum or Bitcoin).
\cite{2019:iot-access} is another system which equips IoT devices with a TEE to guarantee integrity and confidentiality of the IoT data.

\paragraph*{Confidential computation}
BeeKeeper~2.0~\cite{2018:beekeeper} is a blockchain network for IoT systems that consists of IoT devices, servers and validator nodes.
It enables IoT devices to share data with each other.
Furthermore, devices can use servers for performing homomorphic computations on encrypted data.
Computation results are verified by the validator nodes and recorded on the blockchain after successful verification.
Since homomorphic encryption is used, the confidentiality of data sent to and processed by servers is guaranteed.
BeeKeeper~2.0 can be added on top of any blockchain technology (Hyperledger Fabric, Ethereum etc.).
In the paper, authors use Hyperledger Fabric for deployment.

\paragraph*{ARM TZ and blockchain technology}
Some related systems use ARM TZ not for smart contracts directly but in context of the blockchain technology (which is the underlying technology of smart contracts).
Secure blockchain lightweight wallets (SBLWT)~\cite{18:sblwt} use ARM TZ to guarantee confidentiality and integrity for information generated and stored in the Bitcoin wallet (wallet's private key, wallet addresses, block headers used for Simplified Payment Verification).
Synchronization of block headers and the verification process of transactions is executed in the secure world to avoid any manipulation by an attacker.
SBLWT is safer than often used software wallets but still more portable than hardware wallets.
An implementation of SBLWT using ARM TZ with OP-TEE has been deployed on Raspberry Pi 3 Model B.
The Bitcoin wallet~\cite{17:bitcoin-wallet} and sensitive Bitcoin wallet information are backed by ARM TZ technology.
\section{Conclusion and Open Challenges}
\label{sec:conc}

This work reports on our practical experience with the combination of Hyperledger Fabric and ARM TZ, a particularly useful setting in the context of IoT.
We presented the design and implementation of a prototype, \sys, for Hyperledger Fabric chaincode execution with ARM TZ and OP-TEE, demonstrating a fully-working mechanism to execute smart contracts inside the secure world. %
Our design minimizes the TCB executed by avoiding execution of a whole Hyperledger Fabric node inside the TEE, which is assumed to be running in an untrusted environment.
Instead, we restrict it to the execution of only the smart contract.
The \sys prototype exploits the open source OP-TEE framework, as it supports deployments on cheap low-end devices (\emph{e.g.}, Raspberry Pis).
However, we also report on several challenges faced while building our prototype, in particular the missing support for remote attestation in ARM TZ and OP-TEE, and missing hardware support for critical security features. 
Also specifications, such as GP's TEE APIs, can limit the potential of TEEs.
For example, according to the specification a TEE is only capable of doing integer-based computation, therefore limiting the scope of smart contracts that could be executed inside a TEE.
For this reason it is important to reevaluate and update such specifications regularly.
Furthermore, support for a validation mechanism in the blockchain which ensures that the smart contract has been executed correctly and has not been tampered with can be of use.

Our experimental results highlight the performance and energy trade-off due to additional security guarantees provided by ARM TZ.
We leave the final design and replication of chaincodes in our prototype as future work.
The \sys is released as open source.
 
\section*{Acknowledgments}
The research leading to these results has received funding from the European Union's Horizon 2020 research and innovation programme under the LEGaTO Project (\href{https://legato-project.eu/}{legato-project.eu}), grant agreement No~780681.

{\small   
\bibliographystyle{IEEEtranN}
\bibliography{biblio}
}
\typeout{get arXiv to do 4 passes: Label(s) may have changed. Rerun}
\end{document}